\documentclass[manuscript]{aastex}

\shorttitle{Slow waves and fast flows in loops}
\shortauthors{Ofman, Wang, Davila}

\begin{document}


\title{Slow magnetosonic waves and fast flows in active region loops}


\author{L. Ofman\altaffilmark{1,2,3}, T.J. Wang\altaffilmark{1,2}, J.M. Davila\altaffilmark{2}}


\altaffiltext{1}{Catholic University of America, Washington, DC
20064} \altaffiltext{2}{NASA Goddard Space Flight Center, Code 671,
Greenbelt, MD 20771} \altaffiltext{3}{Visiting Associate Professor, Department of
Geophysics and Planetary Sciences, Tel Aviv University, Tel Aviv 69978,
Israel}


\begin{abstract}
Recent EUV spectroscopic observations indicate that slow magnetosonic waves are present in active region (AR) loops. Some of the spectral data were also interpreted as evidence of fast ($\sim100-300$ km s$^{-1}$) quasi-periodic flows. We have performed three-dimensional magnetohydrodynamic (3D MHD) modeling of a bipolar AR that contains impulsively generated waves and flows in coronal loops. The model AR is initiated with a dipole magnetic field and gravitationally stratified density, with an upflow driven steadily or periodically in localized regions at the footpoints of magnetic loops. The resulting flows along the magnetic field lines of the AR produce higher density loops compared to the surrounding plasma by injection of material into the flux-tubes and the establishment of siphon flow. We find that the impulsive onset of flows with subsonic speeds result in the excitation of damped slow magnetosonic waves that propagate along the loops and coupled nonlinearly driven fast mode waves. The phase speed of the slow magnetosonic waves is close to the coronal sound speed. When the amplitude of the driving pulses is increased we find that slow shock-like wave trains are produced. When the upflows are driven periodically, undamped oscillations are produced with periods determined by the periodicity of the upflows. Based on the results of the 3D MHD model we suggest that the observed slow magnetosonic waves and persistent upflows may be produced by the same impulsive events at the bases of ARs.

\end{abstract}



\section{Introduction}
Observations of slow mode oscillations in hot loops associated with flows were seen in X-ray emission in the past \citep[e.g.,][]{Sve94}. Observations with the Solar and Heliospheric Observatory (SOHO) Solar Ultraviolet Measurements of Emitted Radiation (SUMER) instrument \citep[e.g.,][]{Kli02,Wan02,Wan03a,Wan03b} show that standing slow magnetosonic waves are excited in hot ($\sim6-9$MK)  coronal loops by impulsive flare-like phenomena. Recent Hinode satellite Solar Optical Telescope (SOT) observations, and Transition Region and Coronal Explorer (TRACE) satellite Extreme Ultraviolet (EUV) imaging observations show that some coronal loop oscillations are associated with flows and jets in cool ($\sim1$MK) coronal loops  \citep{Asc02,OW08}. The difficulties in distinguishing between slow magnetosonic waves and quasi-periodic flows in some EUV time-sequence images of coronal loops are well known \citep[e.g.,][]{BC99}. The use of spectral data in addition to stereoscopic EUV imaging to help resolve whether the observed event is slow magnetosonic wave or quasi-periodic flow is a promising approach \citep[e.g.,][]{Mar09,MW09}.  However, the quasi-periodic signatures propagating along the loops seen in EUV difference images and quasi-periodic Doppler blue shift in spectral emission lines are still being debated regarding their wave/flow nature due to the limited quality
of the present data: e.g., insufficient spectral and spatial-temporal resolutions
and small signal-to-noise (S/N) ratios \citep{DeP10,Verw10,Wan09,Wan10,Tia11}.

Linear MHD models of slow magnetosonic waves in coronal loops with steady background flows developed in the past were limited to straight cylindrical structures, and linear waves \citep[e.g.,][]{Nak95}. Impulsive excitation of slow magnetosonic waves without flows in bipolar magnetic structures was recently considered in 2D MHD \citep{Sel05b,Sel07}, and recently in 3D MHD \citep{OS09,SO09} models. In these studies it was found that the simultaneously excited fast magnetosonic waves could facilitate the establishment of standing slow magnetosonic waves in curved magnetic field geometry. Recently, \citet{SmO11} studied slow magnetosonic waves in an expanding coronal loop associated with a Coronal Mass Ejection (CME) using 3D MHD. The propagation and dissipations of slow magnetosonic waves in coronal plumes with background solar wind flow were studied using 2.5D MHD \citep{ONS00}. Oscillations of multiple cylindrical coronal loop strands that include coupled slow waves,  transverse waves and the nonlinear interaction with background steady flow were studied using 3D MHD model \citep{Ofm09}.

The present study expands previous models by including for the first time impulsive and non-steady flow injection that excites the waves in magnetic loops of a bi-polar active region (AR) in the full 3D MHD model. The nonlinear interactions of the oscillations and flows are studied in realistic AR geometry in hot coronal AR  plasma model stratified by solar gravity. The results demonstrate that injection of steady and non-steady flows at the footpoints of coronal loops excite compressional waves in longitudinal and transverse directions to the loop's magnetic field. The model shows that the same impulsive process at the corona-transition region boundary can produce flows and waves in coronal AR loops.

\section{Observational Motivation}
\label{obs:sec}
It was suggested that the standing slow-mode waves in hot coronal loops are excited by small (or micro-)
flares at one footpoint \citep{Wan03b,Wan05, Wan11b}. \citet{Wan05} examined the evolution of
Fe\,{\sc{xix}} and Fe\,{\sc{xxi}} line profiles in the initial phase for 54 oscillations, and found
that nearly half show the presence of two spectral components: the stationary component, and the shifted component that reaches maximal Doppler velocity on the order of 100$-$300 km~s$^{-1}$. These features suggest that the
initiation of waves is closely associated with hot plasma upflows produced in flares. 

The oscillation event, occurring at 11:20 UT on 2002 April 17 above the north-west limb, is such a clear example (see Figures~1--4 in \citet{Wan05}). This event was first studied by \citet{Wan05} using SOHO/SUMER observations, and was further analyzed by \citet{Wan11b} using Reuven Ramaty High Energy Solar Spectroscopic Imager (RHESSI) data to explore the trigger mechanism. The SUMER observed two brightenings in the Fe xix line located at the intersections of the slit with a coronal loop. The evolution of Fe xix spectral profiles shows the emergence of a second component, which has the Doppler velocity up to 230 km s$^{-1}$, measured by double Gaussian fits. The slow-mode waves were seen excited following this initial hot flow injection. The RHESSI observed a hard X-ray source near the northern footpoint of the oscillating loop, suggesting that a hot flow pulse produced by a microflare (e.g., via the explosive chromospheric evaporation) moves upward along the loop, leading to the two brightenings as observed by SUMER. This example indicates that the slow-mode wave may be driven by the hot flow that is produced by energy release near the loop's footpoint.

Similarly, the association of continuous upflows and propagating slow waves observed in coronal loops
near the edge of ARs by Hinode/EIS suggests that they may result from the same mechanism,
associated with tiny nanoflare-like small-scale energy release \citep{Wan10, Nis11}. The SUMER oscillation events
have high recurrence rate (2--3 times within two hours), consistent with the hypothesis of their microflaring origin. We speculate that the nanoflares could
produce nearly continuous upflows near the loop's footpoints because of their much higher
recurrence rate,  while larger, lower-frequency flares are responsible for the
discernible quasi-periodic wave disturbances that propagate upwards more than 70 Mm
above the footpoints.  A noticeable difference is detected between the upflow related phenomena observed by EIS and SUMER.
The EIS continuous upflows and propagating slow waves are most clearly observed in warm (1--2 MK)
coronal loops, while those detected by SUMER are observed in hot ($>$6 MK) flaring loops. This observational
difference can be understood by recent nanoflare models \citep[see the review by][]{kli09} that
predict the presence of small amount of very hot ($>$ 5 MK) plasma. There are two main reasons that make hot spectral lines hard to detect: the shorter lifetime and the lower densities of the hot material compared the the warm plasma. Therefore, we model these phenomena
in hot coronal loops as a first reasonable step with the aim of understanding the relationship between flows and waves excitations by impulsive events at the loops' footpoints. 

\section{Numerical Model}
\begin{figure}
\centerline{\includegraphics[scale=0.65]{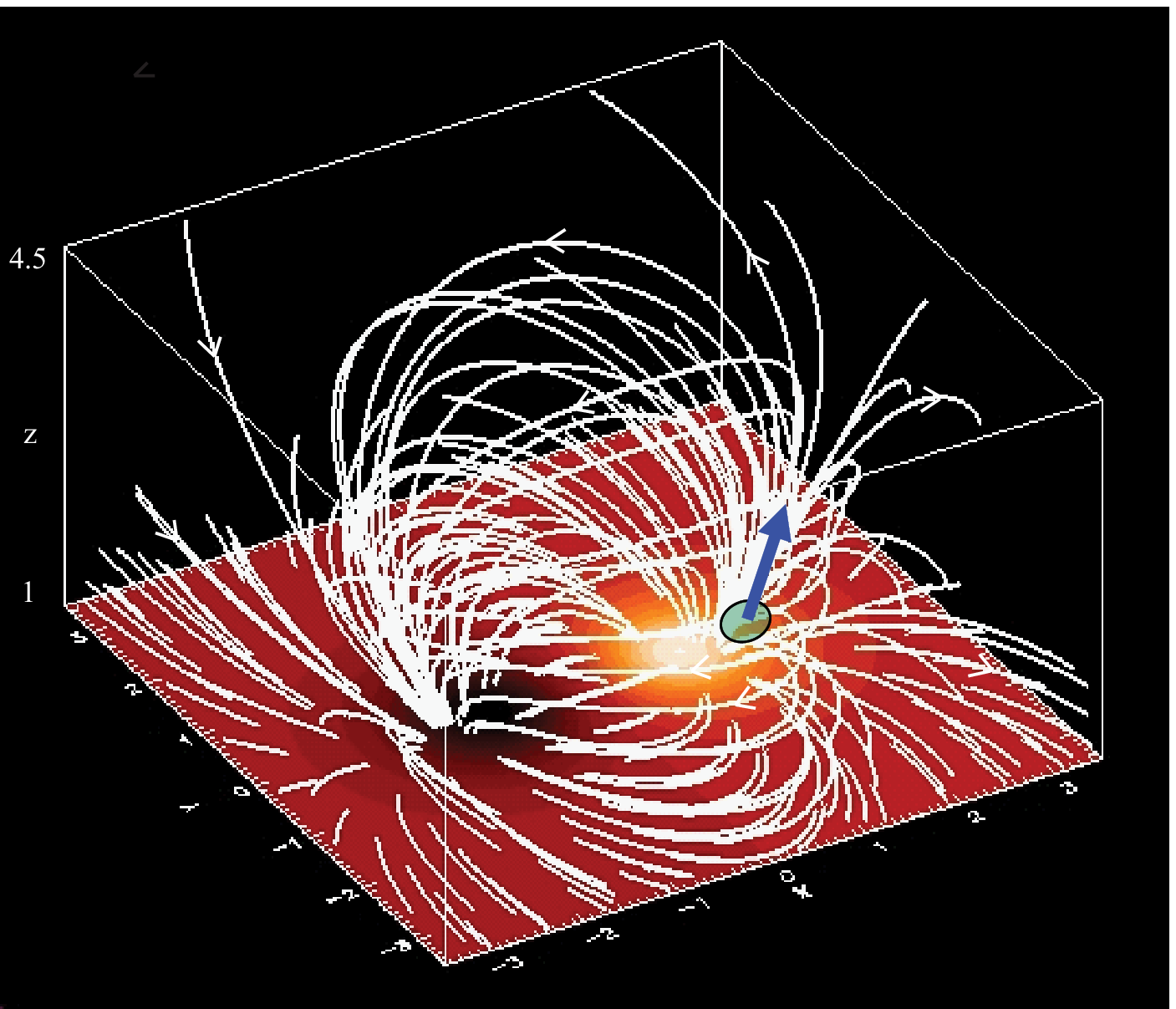}}
\caption{The initial dipole magnetic field (white curves) used for the model AR. The intensity scale shows the magnetic field magnitude at the base of the AR and the white arrows on some field lines show the direction of the magnetic field. The dimensions are in normalized units. The arrow shows the location and direction of the injected flow. }
\label{dipole:fig}
\end{figure}

Here, we describe the numerical 3D MHD model used to study the hot AR loops with flows and waves. The resistive 3D MHD equations are solved with gravity and isothermal energy equation on a Cartesian $258^3$ grid using the modified Lax-Wendroff method with 4th order stabilization (smoothing) term. The initial magnetic field is a dipole (see Figure~\ref{dipole:fig}) with gravitationally stratified normalized density given by
\begin{eqnarray}
\rho=\rho_0 e^{[1/(10+z-z_{min})-0.1]/H},
\end{eqnarray} where $H=2k_BT_0R_s/(10GM_sm_p)$ is the normalized gravitational scale height, $R_s$ is the solar radius, $k_B$ is Boltzmann's constant,
$T_0$ the background (isothermal) temperature, $G$ is the universal gravitational constant, $M_s$ is the solar mass, and $m_p$ is the proton mass, and $\rho_0=1$. The following normalization parameters were used in the present study: distances in units of $a=R_s/10$, magnetic filed $B_0=100$ G, $T_0=6.3$ MK (to model the hot coronal AR loops), number density $n_0=1.38\times10^9$ cm $^{-3}$, resulting in Alfv\'{e}n speed $V_{A0}=B_0/(4\pi n_0)^{1/2}=5872$ km s$^{-1}$ used to normalize the velocities, the Alfv\'{e}n time $\tau_A=a/V_{A0}=11.9$ s used for time normalization, the isothermal sound speed $C_s=323$ km s$^{-1}=0.055V_{A0}$, and the corresponding plasma $\beta_0=\frac{2C_s^2}{V_{A0}^2}=0.00605$. The model is based on the 3D MHD model of a bipolar AR developed initially by \citet{OT02}. This model and its extensions and variants were used successfully in a number of studies of waves in ARs and in coronal loops \citep[e.g.,][]{Ofm07,MO08,SO09,SeO10,SOS11,SSO11}. As in previous studies we set the Lundquist number to $S=10^4$, i.e., the resistive diffusion time scale is 4 orders of magnitude longer than the Alfv\'{e}n time. Since we are dealing with time scales $O(100\tau_A)$ - the resistive dissipation is insignificant. Numerical convergence study has shown that the results reported below are not affected by numerical dissipation. 

\begin{figure}
\centerline{\includegraphics[scale=0.58]{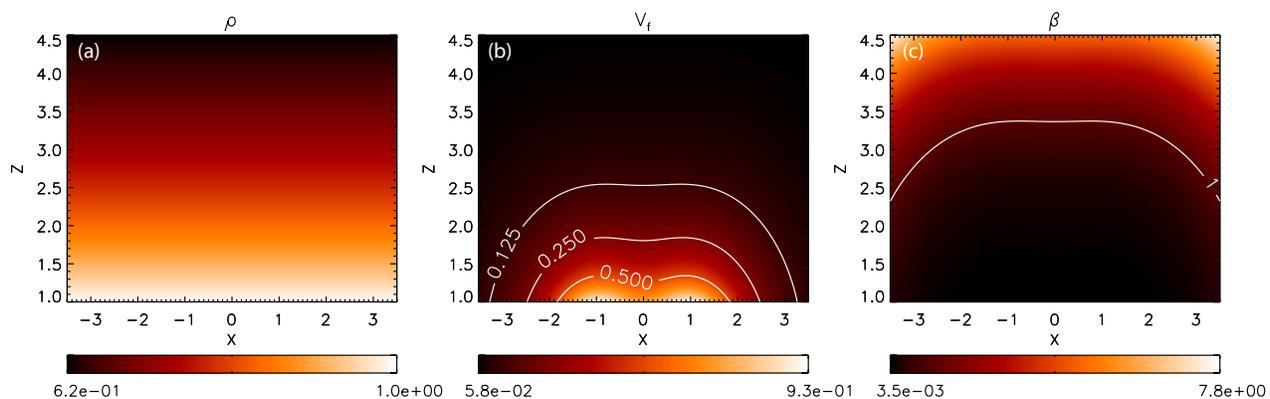}}
\caption{The initial density (left), fast magnetosonic speed (middle), and plasma $\beta$ (right) in the $xz$ plane at $t=0$ in the model AR. The contours on $V_f$ show the 50\%, 25\%, and 12.5\% levels of the maximal value, and the contour $\beta=1$ shows the heights where this value is reached. }
\label{Vf_beta_rho_xz_t0:fig}
\end{figure}

In Figure~\ref{Vf_beta_rho_xz_t0:fig} the initial density, $\rho$, the fast magnetosonic speed $V_f=(V_A^2+C_s^2)^{1/2}$, where $V_A=B(x,y,z)/\rho(x,y,z)^{1/2}V_{A0}$ is the local Alfv\'{e}n speed, and the plasma $\beta$ in the $xz$ plane at $y=0$ of the model AR are shown. The gravitational stratification of the density is evident, and the increase of $V_f$ in the lower central region due to the bipolar magnetic field of the AR is apparent.  The plasma $\beta$ increases with height due to the rapid decrease of the dipole magnetic pressure $\sim B^2$ compared to the decrease of the thermal pressure with height.

Nonsteady and impulsive flow injection along the field is introduced at the lower boundary of the model AR as
\begin{eqnarray}
&&\mbox{\bf V}=V_0(x,y,z=z_{min},t)\mbox{\bf B}/|B|,
\end{eqnarray} where
\begin{eqnarray}
&&V_0(x,y,z=z_{min},t)=A_v(t)V_A exp\left\{-\left[\left(\frac{x-x_0}{w_0}\right)^2+\left(\frac{y-y_0}{w_0}\right)^2\right]^2\right\},
\label{v0:eq}\end{eqnarray} with
\begin{eqnarray}
&&A_v(t)=A_0,\ t_0<t<t_{max},
\end{eqnarray} for impulsively excited continuous flow or
\begin{eqnarray}
&&A_v(t)=A_0(1-cos\omega t)/2,\ t>t_0
\end{eqnarray} for periodic upflow. Here, the parameters are $z_{min}=1$, $x_0=0.8$ for the short magnetic loop (length $=2.13$), $x_0=1.2$ for the long magnetic loop (length $=4.53$), $y_0=0$, $w_0=0.12$, $\omega=0.21$, $t_0=0$, and the values $A_0=0.01$, and $0.05$ that corresponds to subsonic upflow were used. The additional power of two on the RHS of Equation~(\ref{v0:eq}) models an upflow with sharper than Gaussian cross-sectional profile. The upflow is imposed at the lower boundary $z=z_{min}=1$ at a circular area with twice the radius $2w_0$ centered at the right footpoint of the loops. Downflows through the lower boundary are allowed at the other footpoint of the loops. The rest of the boundary conditions at $z=1$ are: $v_x=v_y= 0$ outside the upflow/downflow regions, with fixed $\mbox{\bf B}$,    density $\rho$ and $v_z$ extrapolated from the interior points. These boundary conditions approximate the line-tied boundary conditions for coronal loops \citep[e.g.,][]{Lio98,MO08}, modified allowing inflow and outflow. The line-tied boundary conditions approximate the effects of the dense photosphere on the loop's magnetic footpoints, and ensure wave reflection necessary for establishment of a standing wave with nodes at the footpoints. Open boundary conditions are used at the top and side planes of the computational box allowing wave propagation outward through the boundary with negligible reflection.

The magnetic field lines of the loops and the dependence of the loops parameters on the coordinate along each loop, $s$, are shown in Figure~\ref{loop_s:fig}. It is interesting to note that the normalized magnetic field strength, $B$ and the fast magnetosnic speed $V_f$ along the loop decrease with height by a factor of 3 for the short loop and by an order of magnitude for the long loop. The significant variation of the magnetic field and the fast magnetosonic speed is usually neglected in linear coronal seismology models \citep[see the review by][]{NV05}. The density along the loops varies by $\sim10\%$ for the short loop, and $\sim23\%$ for the long loop. The small variation of the density with height (compared to cool 1MK loops) is the result of the high temperature (6.3MK) and the low height of these loops.  

\section{Numerical Results}
In Figures~\ref{rh_v_xz_sl:fig}-\ref{rh3d:fig} the results of the 3D MHD model of steady and periodic  inflows and associated waves are shown for the initial state and parameters described in the previous section. The evolution of the flows and oscillation is also evident in the animations accompanying the online version of the journal. First, we report the results for the steady inflow cases. In Figure~\ref{rh_v_xz_sl:fig} the  snapshots of the density and the velocity in the $xz$ plane at the end of the run ($t=280\tau_A$) of steady inflow with $V_0=0.05$ are shown. The intensity scale indicates the magnitudes of the density and the velocity, and the arrows in the velocity images indicate the direction of the flow for velocities greater than 6\% of the maximal value. The short loop results are shown in Figure~\ref{rh_v_xz_sl:fig}a-b, and the long loop results are in Figure~\ref{rh_v_xz_sl:fig}c-d. It is evident that the steady inflow leads to the formation of a steady short loop in the low-$\beta$ region of the model AR. The short loop is filled with inflowing coronal plasma and reaches higher (nonuniform) density than the surrounding plasma by a factor $\sim1.5$. Since the EUV line emission in optically thin coronal plasma is proportional to $n^2$ this loop would appear twice brighter than the same loop preceding the inflow in background-subtracted  EUV images. In Figure~\ref{rh_v_xy_s:fig} the cut in the $xy$ plane at $z=1.26$ of the density  and the velocity of the AR with the short loop are shown. The locations of the upflow and downflow that correspond to the loop with higher than the surroundings density are evident. In addition the circular structures in the density and the velocity are formed due to the compression and rarefaction produced by the flow in the AR and the fast magnetosonic waves emitted in the initial impulsive stage of the loop formation.

The time-distance plots shown in Figure~\ref{td_sl:fig} for the boxed regions E-H, (Figure~\ref{rh_v_xz_sl:fig})  demonstrate the details of the loops oscillations. The widths of the boxes are 7 gird points in space and the time-distance plots are smoothed with 15 points running average. The snapshots of the boxes are taken every 5$\tau_A$. The time-distance plot E is aligned along the $z$ direction centered at $x=0$ through the apex of the short loop. It is evident that the temporal oscillation of the loop in the $z$ direction is small, and the density variations are due primarily to compressive density oscillation, following an initial transient stage (the time for the inflow to reach the apex of the loop). The time-distance plot F is aligned along the $x$ direction centered at height $z=1.55$ through the apex of the loop and shows the oscillatory displacement of the loop top, with evidence of compressive density oscillations. By comparing time-distance plots E and F it is evident that the dominant transverse (kink) oscillation of the loop is in the $x$ direction (see, Figure~\ref{vbn_t_siphon_v0_05xp00_8_a:fig} below).  The time-distance plots G (centered at $z=1.27$) and H (centered at $z=1.82$), are cuts in the $x$ direction through the footpoints of the short and long loops, respectively.  Note, that the positive values of $x$ correspond to the right side of the plots in Figure~\ref{rh_v_xz_sl:fig}. Here, the transverse oscillations of the footpoints, as well as the effects of density compression are evident. The displacements of the 
loops' two footpoints are in-phase, i.e., the loop sways left and right in the loops' curvature plane, 
whereas the intensity oscillations near the two footpoints are in anti-phase, a signature of the fundamental 
slow mode wave, which has the density anti-nodes at the footpoints. These loop oscillations are similar to the second harmonic of the vertical kink mode discussed by \citet{Wan08} in connection with TRACE EUV coronal active region loop oscillations. 

\begin{figure}
\centerline{\includegraphics[scale=0.9]{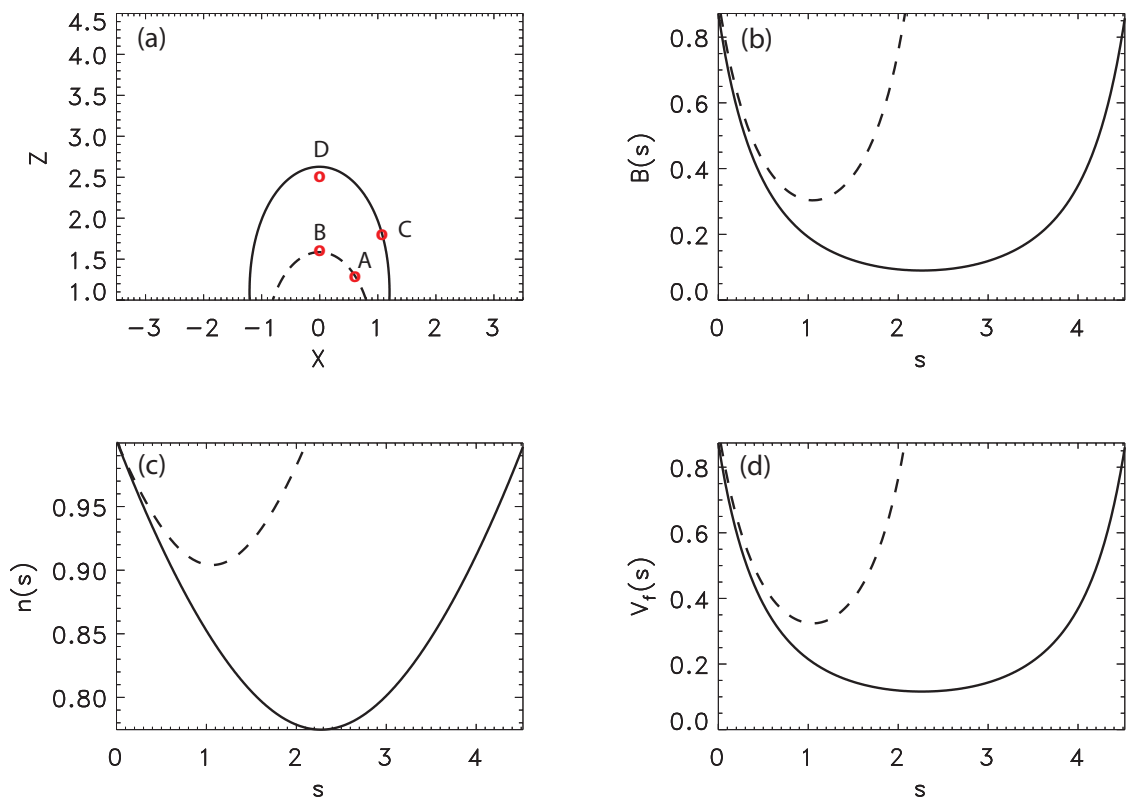}}
\caption{The initial structure of the long (solid line style) and short (dashes) loops. (a) The magnetic field lines at the center of the loops in the $xz$ plane at $y=0$. The red circles A, B, C, D mark the locations of temporal evolution plots of the variables shown below. (b) The magnetic field intensity along the loops, where $s$ is the coordinate along each loop. (c) The density along the loops. (d) The fast magnetosonic speed, $V_f(s)$ along the loops.}
\label{loop_s:fig}
\end{figure}

\begin{figure}
\centerline{\includegraphics[scale=0.85]{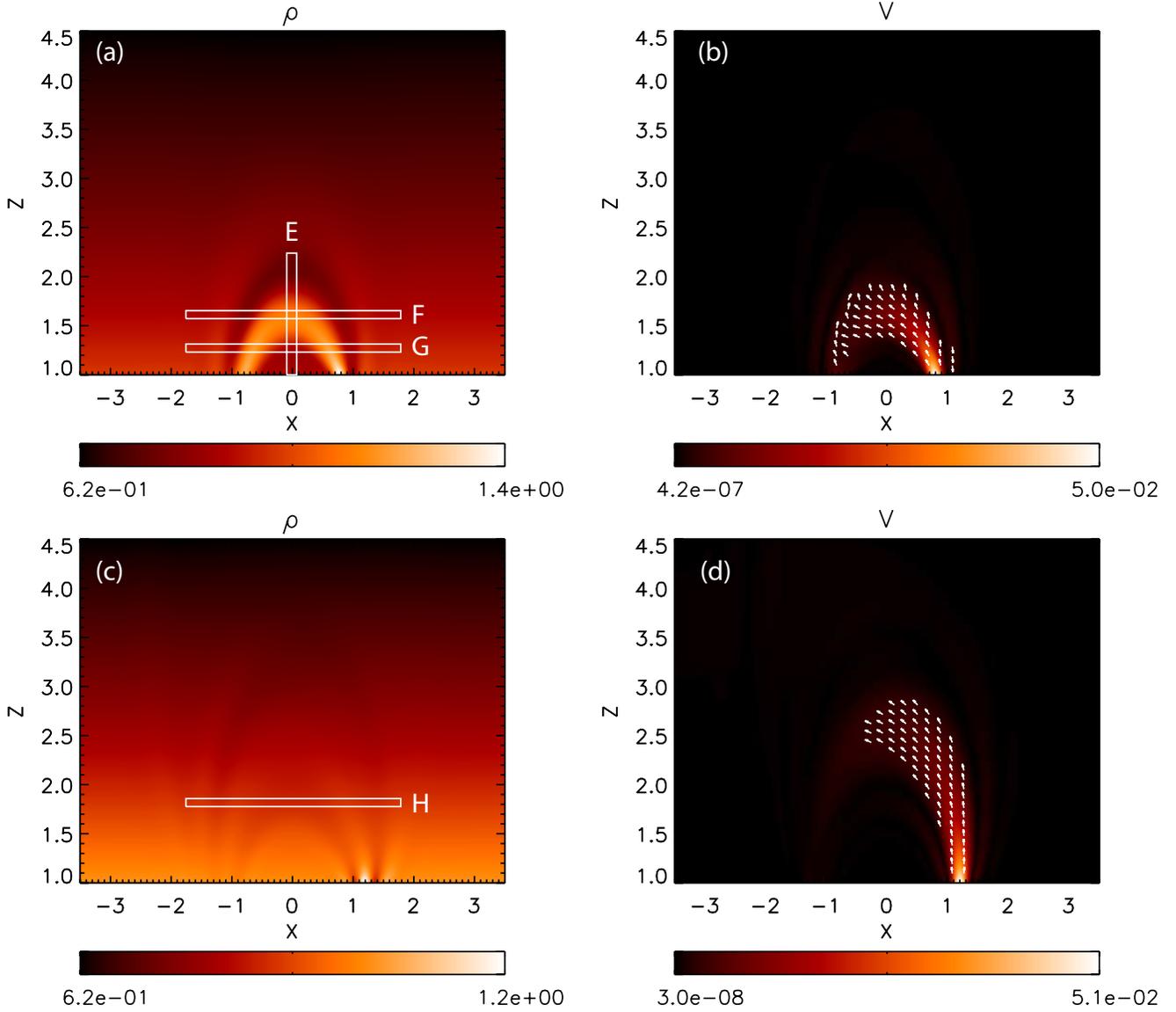}}
\caption{The snapshots of the density and the velocity in the $xz$ plane at $t=280\tau_A$ with steady inflow ($V_0=0.05$). The intensity scale shows the magnitude of the density and the velocity, and the arrows indicate the direction of the flow for velocities greater than 6\% of the maximal value. (a) Density for the short loop; (b) velocity in the short loop; (c) density for the long loop; (d) velocity in the long loop. The rectangles marked E-H mark the locations of the time-distance plots shown in Figure~\ref{td_sl:fig}. Animations of the density are available in the online version of the journal.}
\label{rh_v_xz_sl:fig}
\end{figure}

\begin{figure}
\centerline{\includegraphics[scale=0.74]{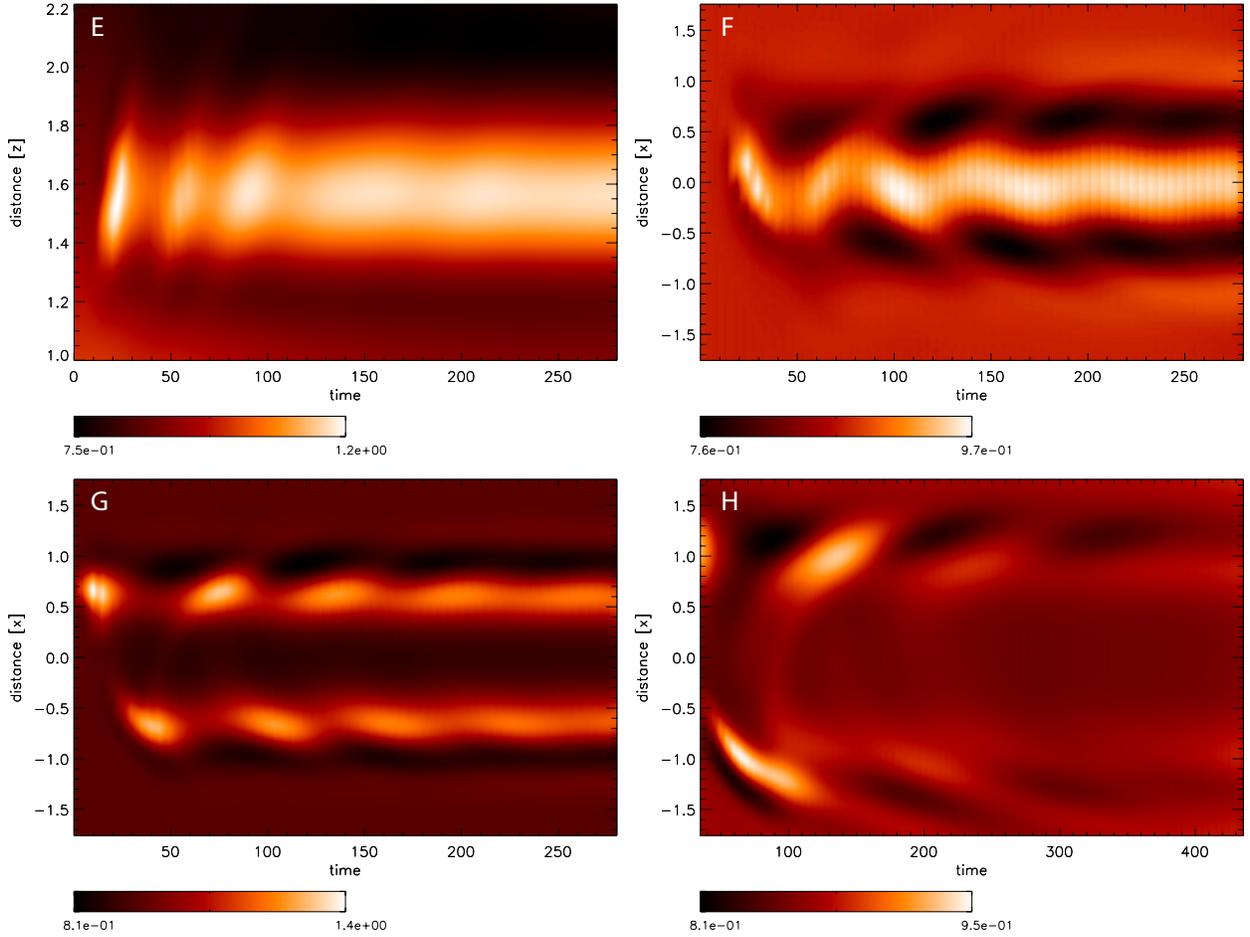}}
\caption{The time-distance plots for the boxed regions E-G of the short loop and H of the long loop shown in Figure~\ref{rh_v_xz_sl:fig}. Panel E is the time-distance plot obtained from the cut E in the $z$ direction of the short loop centered at $x=0$. Panel F is the time-distance plot of the cut F in the $x$-direction centered at $z=1.82$. Panel G is the time-distance plot of the cut G in the $x$-direction centered at $z=1.27$ of the short loop. Panel H is the time-distance plot of the cut H in the $x$-direction centered at $z=1.82$. In panel H the initial transient time is removed to show better the subsequent evolution of the loop's density. In all panels the temporal cadence is 5$\tau_A$. Both, displacement and density oscillations are present showing evidence for the transverse and compressional waves in the $x-z$ plane of the loops.}
\label{td_sl:fig}
\end{figure}

\begin{figure}
\centerline{\includegraphics[scale=0.8]{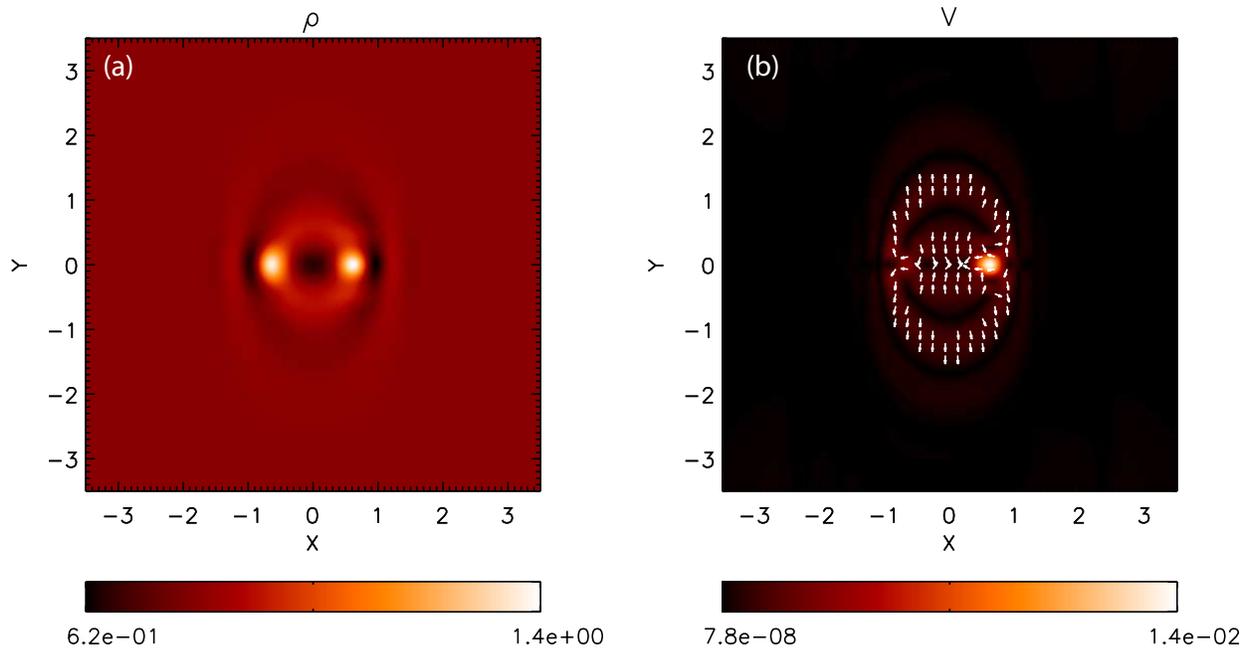}}
\caption{Same as \ref{rh_v_xz_sl:fig}, but in the $xy$ plane for short loop. (a) Density; (b) velocity.  Animation of the density is available in the online version of the journal.}
\label{rh_v_xy_s:fig}
\end{figure}

\begin{figure}
\centerline{\includegraphics[scale=0.8]{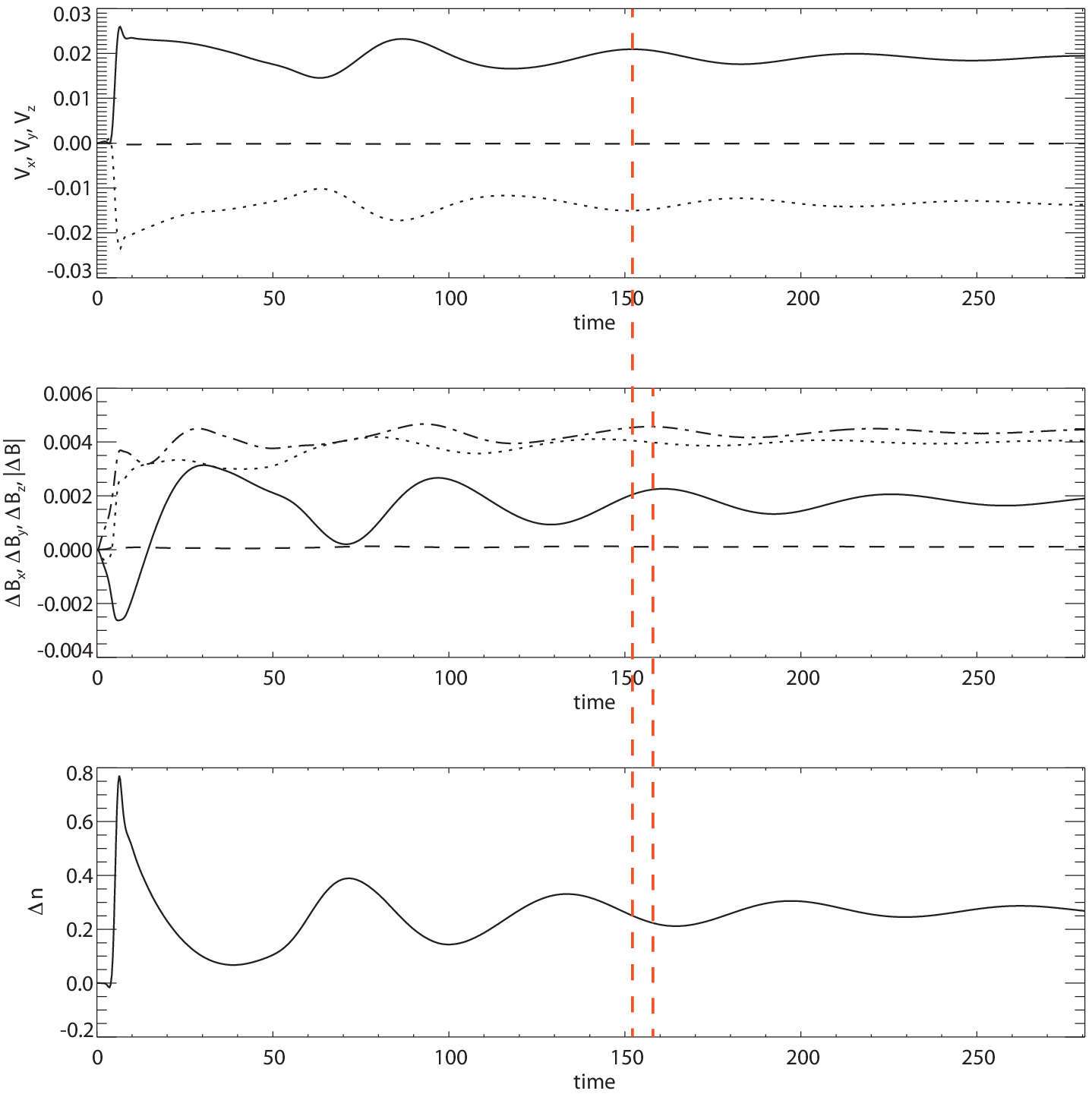}}
\caption{The temporal evolution in a short loop of the velocity components ($V_x$: dots; $V_y$: dashes; $V_z$: solid) (top panel), the perturbed magnetic field components ($B_x$: dots; $B_y$: dashes; $B_z$: solid; $|\Delta B|$: dot-dashes) (middle panel), and the perturbed density (lower panel) at the right footpoint position $A=(0.59,-0.01, 1.27)$ for steady inflow. The red dashed lines  through the third peak of $V_z$ and the peak of $|\Delta B|$ help evaluate the phase shifts between these variables.}
\label{vbn_t_siphon_v0_05xp00_8_f:fig}
\end{figure}

\begin{figure}
\centerline{\includegraphics[scale=0.8]{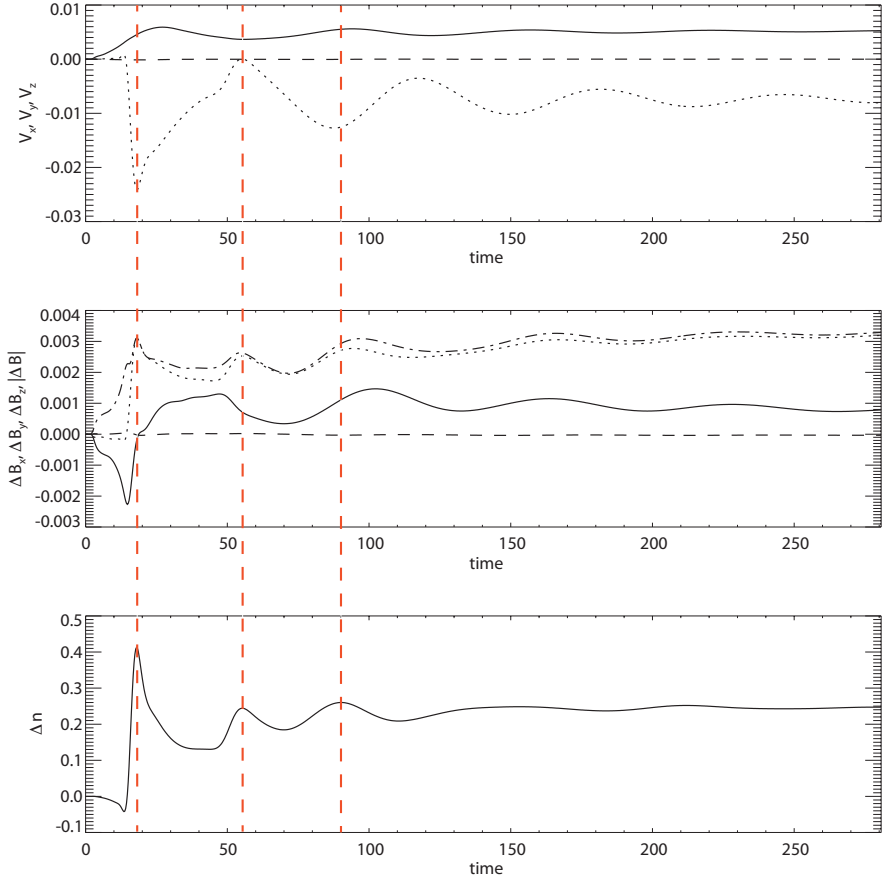}}
\caption{The temporal evolution in a short loop of the velocity components ($V_x$: dots; $V_y$: dashes; $V_z$: solid) (top panel), the perturbed magnetic field components ($B_x$: dots; $B_y$: dashes; $B_z$: solid; $|\Delta B|$: dot-dashes) (middle panel), and the perturbed density (lower panel) at the apex of the loop $B=(0.01,-0.01, 1.6)$ for steady inflow. The red dashed line marks three peaks of $\Delta n $ helping evaluate the phase shift between the variables.}
\label{vbn_t_siphon_v0_05xp00_8_a:fig}
\end{figure}

\begin{figure}
\centerline{\includegraphics[scale=0.8]{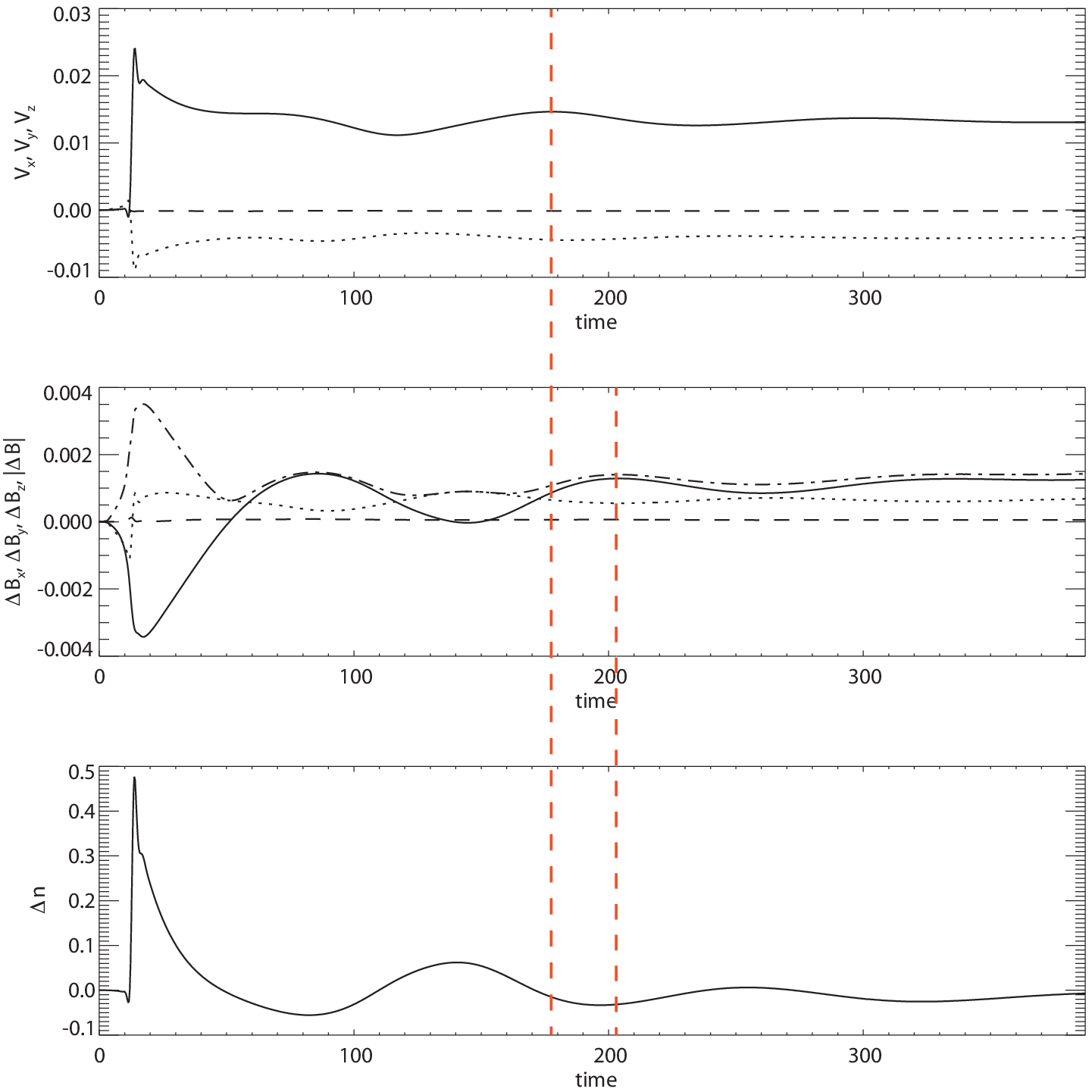}}
\caption{The temporal evolution in a long loop of the velocity components ($V_x$: dots; $V_y$: dashes; $V_z$: solid) (top panel), the perturbed magnetic field components ($B_x$: dots; $B_y$: dashes; $B_z$: solid; $|\Delta B|$: dot-dashes) (middle panel), and the perturbed density (lower panel) at the right footpoint position $C=(1.08,-0.01, 1.82)$ for steady inflow. The red dashed lines  through the third peak of $V_z$ and the peak of $|\Delta B|$ help evaluate the phase shifts between these variables.}
\label{vbn_t_siphon_v0_05xp01_2:fig}
\end{figure}

\begin{figure}
\centerline{\includegraphics[scale=0.58]{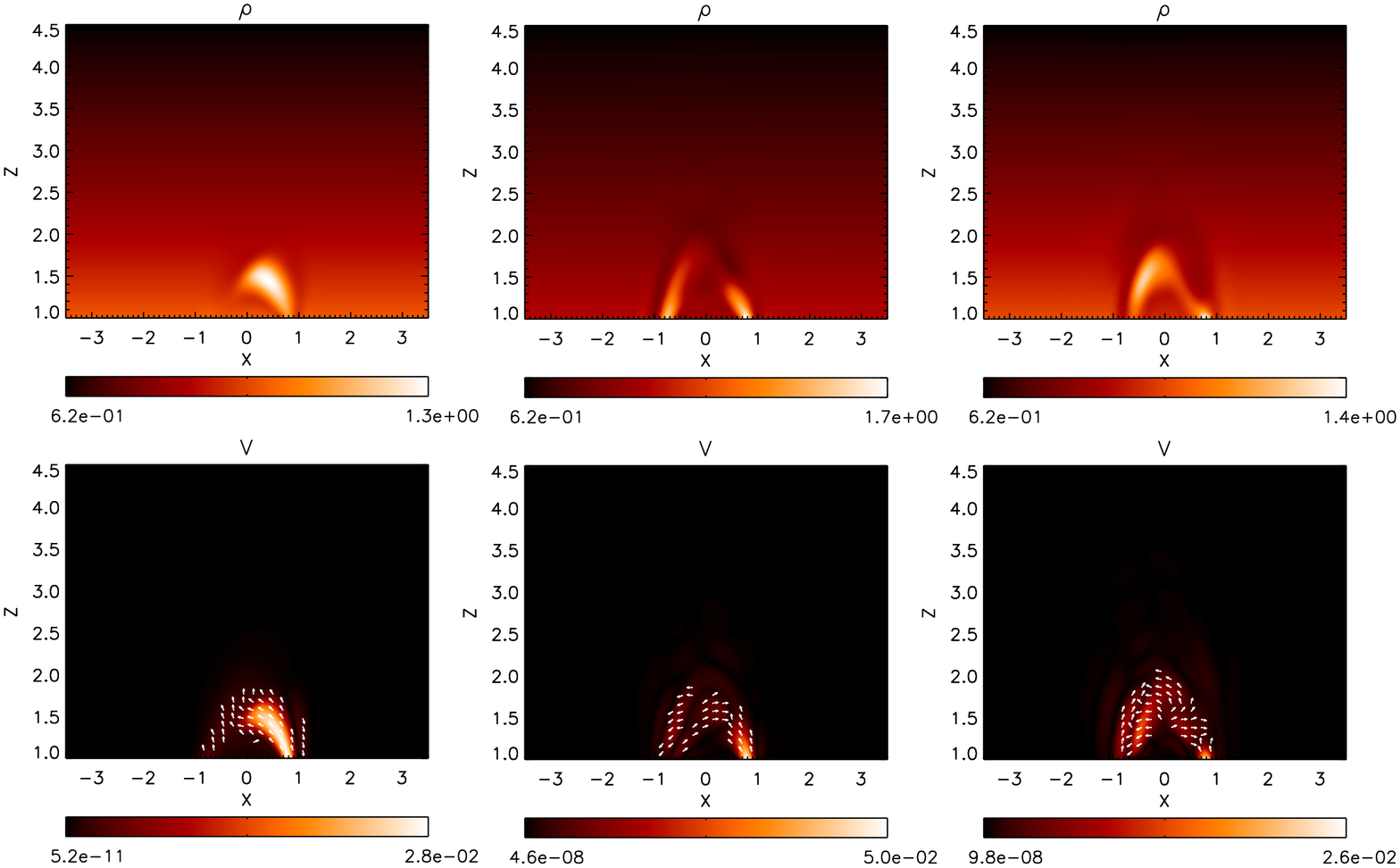}}
\caption{The density (top panels) and the velocity (lower panels) in the $xz$ plane ($y=-0.0136$) at $t=22.5$, 105, 188 $\tau_A$ (from left to right) in a short loop injected periodically with flow. The arrows show the direction of the flow and the intensity shows the magnitude of the flow and the density.  Animations of the density are available in the online version of the journal.}
\label{rh_v_xz_s_periodic:fig}
\end{figure}

\begin{figure}
\centerline{\includegraphics[scale=0.8]{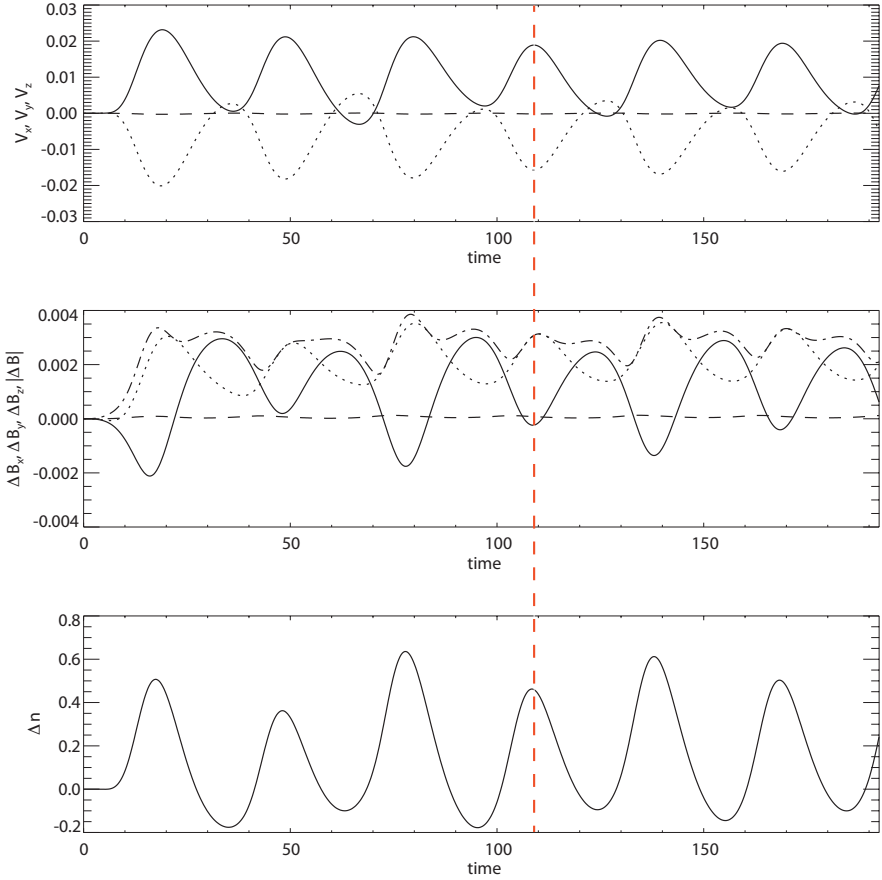}}
\caption{The temporal evolution of the velocity components ($V_x$: dots; $V_y$: dashes; $V_z$: solid), the perturbed magnetic field components ($B_x$: dots; $B_y$: dashes; $B_z$: solid; $|\Delta B|$: dot-dashes), and the perturbed density  at the right footpoint position $A=(0.59,-0.01, 1.27)$ of the short loop with periodic inflow. The red dashed through the 4th peak of $V_z$ line helps evaluate the phase shift between the variables.}
\label{vf30v0_05xp00_8:fig}
\end{figure}

\begin{figure}
\centerline{\includegraphics[scale=0.58]{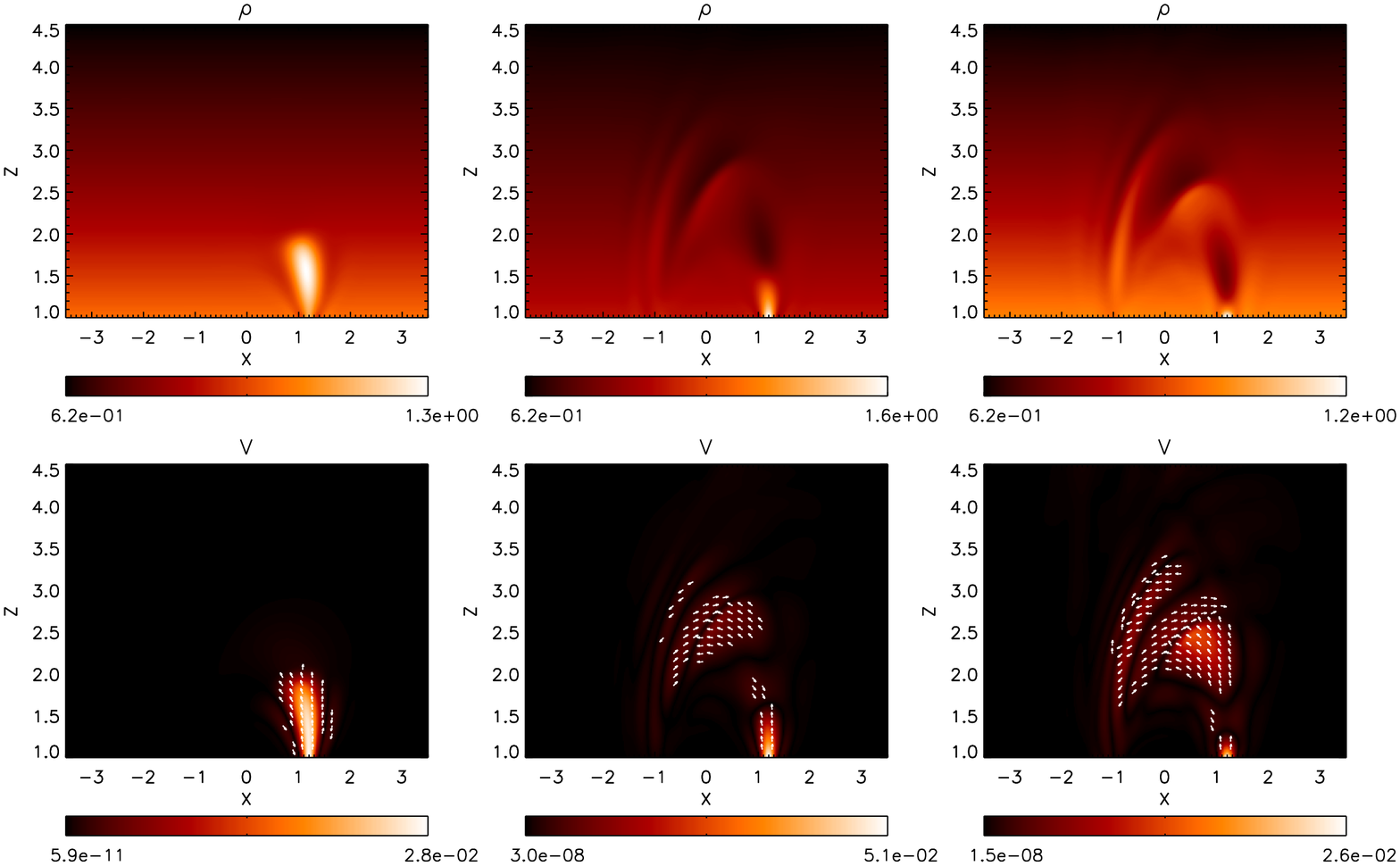}}
\caption{The density (top panels) and the velocity (lower panels) in the $xz$ plane at $t=22.5$, 105, 188 $\tau_A$ (from left to right) in a long loop injected periodically with flow. The arrows show the direction of the flow and the intensity shows the magnitude of the flow and the density.  Animations of the density are available in the online version of the journal.}
\label{rh_v_xz_l_periodic:fig}
\end{figure}

\begin{figure}
\centerline{\includegraphics[scale=0.8]{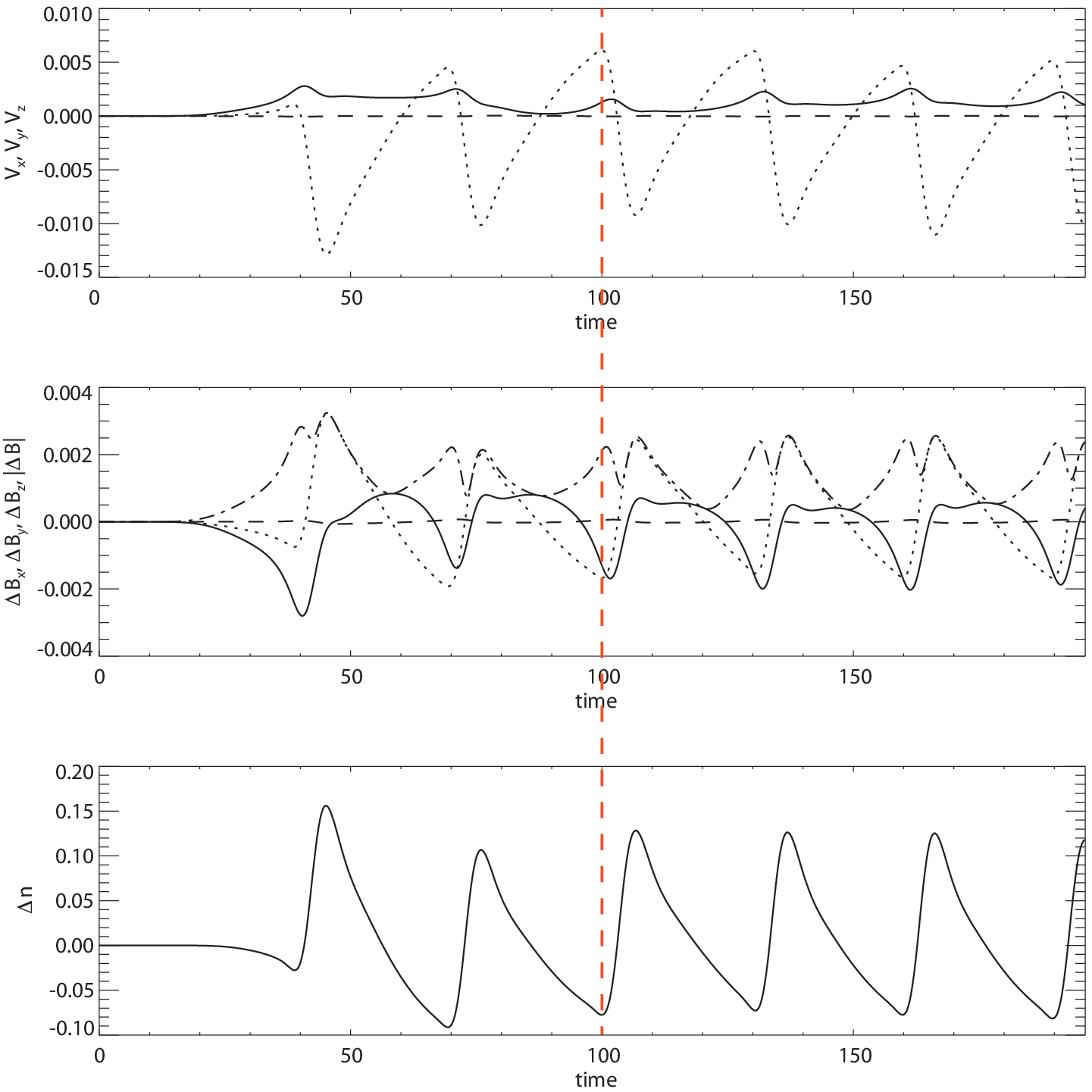}}
\caption{The temporal evolution of the velocity components ($V_x$: dots; $V_y$: dashes; $V_z$: solid), the perturbed magnetic field components ($B_x$: dots; $B_y$: dashes; $B_z$: solid; $|\Delta B|$: dot-dashes), and the perturbed density near the apex of the loop at position $D=(0.01,-0.01, 2.5)$ of the long loop with periodic inflow. The red dashed line though the 3rd peak of $V_x$ helps evaluate the phase shift between the variables.}
\label{vf30v0_05xp01_2:fig}
\end{figure}

\begin{figure}
\centerline{\includegraphics[scale=0.7]{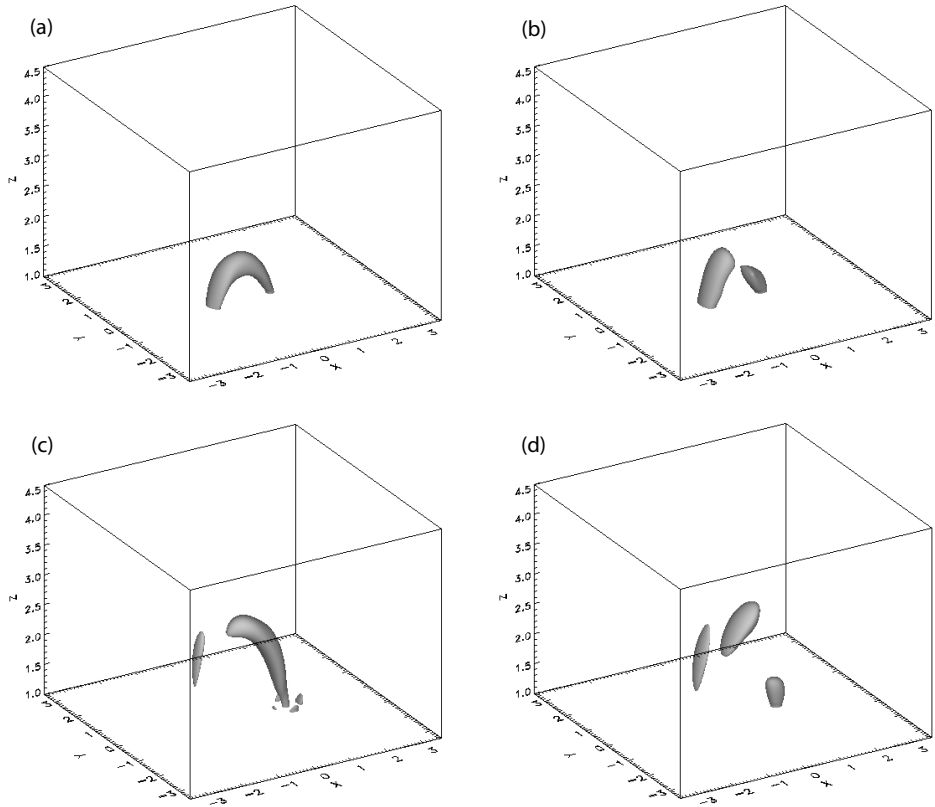}}
\caption{Perturbed density isocontours due to inflow in the short loop: (a) steady inflow, isocontour level at $\rho=0.12$; (b) periodic inflow, isocontour level at $\rho=0.1$. Same for the long loop: (c) steady inflow, isocontour level at $\rho=0.035$; (d) periodic inflow, isocontour level at $\rho=0.05$. The animations of the density isocontours for the four cases are available in the online version of the journal. }
\label{rh3d:fig}
\end{figure}

Figure~\ref{vbn_t_siphon_v0_05xp00_8_f:fig} shows the temporal evolution of the velocity components, the perturbed magnetic field components, and the perturbed density at the  right footpoint of the short loop (point $A=(0.59,-0.01,1.27)$) with steady inflow. At this location the velocity along the loop has two components in the Cartesian frame ($x$ and $z$) as evident in Figure~\ref{loop_s:fig}a. It is evident that it takes about 15$\tau_A$ for the perturbation along the loop to reach this position. The loop contains both, compressive and transverse oscillations, as evident in the time dependence of the velocity and magnetic field  component oscillations, as well as the density fluctuations. The transverse waves are excited by the momentum of the injected velocity pulse and the centrifugal force exerted by the inflow on the curved (dipole) field lines, with the Lorentz force acting as the restoring force. The slow mode wave period is $65\tau_A$ (determined from the last oscillation period of $\Delta n$), and the oscillations damp in several periods due to leakage at the footpoints and outside the loop as a result of  finite $\beta$ allowing oblique propagation and leakage of the slow magnetosonic waves in curved magnetic geometry. The quarter period phase shift between the density and the velocity components along the loop indicate the presence of standing slow-mode waves is established quickly. The presence of the slow mode wave is also evident from nearly anti-phase relation between $|\Delta B|=(\Delta B_x^2+\Delta B_y^2+\Delta B_z^2)^{1/2}$ and  $\Delta n$ (see the vertical red dashes lines in Figure~\ref{vbn_t_siphon_v0_05xp00_8_f:fig}). The anti-phase relation is not exact due to the effects of nonlinearity, and non-modality (i.e., the oscillation are not exact normal modes of the loop) of the oscillations. The effects of nonlinearity are evident in the non-sinusoidality of the oscillations at the initial stages of the evolution. The calculations were repeated with smaller velocity amplitude ($V_0=0.01$) confirming that these effects are due to  nonlinearity and the magnitude of the initial pulse.  

Figure~\ref{vbn_t_siphon_v0_05xp00_8_a:fig} shows the temporal evolution of the velocity components, the perturbed magnetic field components, and the perturbed density at the apex of the short loop (point $B=(0.01,-0.01,1.60)$) with steady inflow. The red dashed line marks three consecutive peaks of the density perturbation, $\Delta n$. It is evident that $\Delta B_x$ is in phase with $\Delta n$ indicating the presence of fast mode wave and the corresponding kink oscillation of the curved loop in the $xz$ plane induced at the impulsive initial stage of the inflow. The variations in $\Delta n$ and $\Delta B_x$ are caused by the oscillations of the nonuniform loop with respect to the fixed point $B$, with contribution from compressive effects (see, Figure~\ref{td_sl:fig}). The period of the oscillation is 35$\tau_A$ based on the second and third peaks. This period is shorter than slow wave period and longer than the kink mode period of the fundamental mode for a cylinder of length $L$ \citep[e.g.,][]{REB84} given by $2L/C_k$, where $C_k=V_A (2/(1+\rho_o/\rho_i)^{1/2}=1.1V_A$, with the ratio of the density inside and outside the loop  $\rho_i/\rho_o=1.25/0.78=1.5$ from the model calculations. Noting that $V_A$ decrease with height in the loop we get $C_k=0.35V_{A0}$ at the apex of the loop. However, the theoretical kink mode period of $12\tau_A$ is a factor of 3 shorter than the period found in the model. Thus, the observed fast mode is not a linear fundamental kink mode of the loop but nonlinearly driven fast mode oscillation with similar properties to the second harmonic of the vertical kink mode. Further justification of this interpretation is obtained by inspecting the phase and period relation between $\Delta B_x$, $\Delta n$, and $\Delta V_x$. It is evident that $\Delta V_x$ period is twice as long as the $\Delta B_x$, $\Delta n$, and peaks, nodes, and minima of $\Delta V_x$ all correspond to peaks of $\Delta n$ and $\Delta B_x$, suggesting that the fast mode is driven nonlinearly by a term that depends on $(\Delta V_x)^2$. Such second order term appears in the $x$ component of the momentum equation $\sim\rho V_x\frac{\partial V_x}{\partial x}=\frac{\rho}{2}\frac{\partial (V_x)^2}{\partial x}$. The nonlinearly driven fast mode waves damp as the driving amplitude of the slow waves decreases with time.

Figure~\ref{vbn_t_siphon_v0_05xp01_2:fig} shows the temporal evolution of the velocity components, magnetic field components, and the perturbed density at the right footpoint of the long loop (point $C=(1.08,-0.01,1.8)$) with steady inflow. The long loop exhibits similar evolution and contains both, longitudinal and fast mode waves induced by the flow. The slow wave period is $\sim 136\tau_A$, determined from the last half period of $\Delta n$ oscillation. The presence of the slow mode wave is also evident from nearly anti-phase relation between $|\Delta B|$ and  $\Delta n$ (see the vertical red dashes lines in Figure~\ref{vbn_t_siphon_v0_05xp01_2:fig}). The anti-phase relation is not exact due to the effects of nonlinearity, and non-modality (i.e., the oscillation are not exact normal modes of the loop) of the oscillations. As in the short loop case the waves damp due to leakage of the slow wave outside the loop. Evidence for the formation of slow standing wave is seen in the nearly quarter period phase shift between the velocity along the loop and $\Delta n$. However, in the case of the long loop the periods of the oscillations is longer as expected,  and the formation of the slow standing wave takes longer time than in the short loop. By comparing the short and long loop oscillations it is evident that the periods of the oscillations and the wavelengths are related to the loop lengths.  The ratio of their periods is about a factor of two, close the loop's length ratio of $2.15$. Wave period values in both loops are in qualitative agreement with the classical expression $2L/C_s$ for the fundamental mode of the slow magnetosonic wave in a straight cylinder \citep[e.g.,][]{REB84} with the short loop in better agreement. However, quantitatively the slow wave periods is smaller by  16\% for the short loop and by 17\% for the long loop case. The difference is mainly due to the curved magnetic geometry and nonuniformity of the loops.

In the following, we report the results for the periodic flow injection cases. The density and velocity distributions in the $xz$ plane (at $y =-0.0136$) cuts of the AR produced by periodic flow injection at the lower boundary at $(0.8,0,1)$ at times $t=22.5$, 105, 188 $\tau_A$ for the short loop are shown in Figure~\ref{rh_v_xz_s_periodic:fig}.  The loop density structure shows time-dependent variation due to the combined effects of the periodic flow injection and the compression of AR plasma by the velocity pulse. The velocity magnitude and the direction vary approximately in-phase with the density showing longitudinal variations along the expanding magnetic field of the loop. The temporal evolution of the velocity components, perturbed magnetic field components, and perturbed density are shown in the short loop at the right footpoint at point $A=(0.59,-0.01,1.27)$ for the periodic flow injection in Figure~\ref{vf30v0_05xp00_8:fig}. It is evident that after the short initial phase as the perturbations reach the above position, the loop's magnetic field, velocity, and density exhibits quasi-periodic oscillations induced by the periodic flow injection at the footpoint. The velocity fluctuations along the loop are in phase with the density perturbation, and in anti-phase with the magnetic field components. The temporal evolution of $|\Delta B|$  shows two dominant frequencies due to the combined effect of the driver and the slow mode wave  in the loop. No temporal damping or growth of the oscillations is evident in the driven case at the steady state, since the fluctuation are injected at the right footpoint continuously, and exit the loop at the left footpoint boundary. The non-sinusoidality of the oscillations is evident and is due to the nonlinear interaction between the flow periodicity, the resulting variation of the local loop Alfv\'en speed that affects the properties of the Alfv\'{e}nic fluctuations, and the waves produced in the loop. The nonlinear effects are reduced when the same calculation is repeated with smaller injected velocity amplitude ($V_0=0.01$, not shown). 

In Figure~\ref{rh_v_xz_l_periodic:fig} the cuts in the $xz$ plane of the AR's density and velocity injected with periodic flow at the lower boundary at $(1.2,0,1)$  at times $t=22.5$, 105, 188 $\tau_A$ for the long loop are shown. It is evident that in this case the flow is divergent as expected in the bipolar magnetic field geometry.  In the case of the long magnetic loop the density structure is more complex than in the case of the short loop that is  entirely in the low-$\beta$ region of the AR as  shown above. Since the long loop reaches higher altitude than the short loop, nonlinear steepening of the  density perturbation fronts are produced due to the decrease of the gravitationally stratified density and the increase of the plasma $\beta$ with height (see Figure~\ref{Vf_beta_rho_xz_t0:fig}). Coupled slow and fast magnetosonic shock-like waves are produced as evident in the compression fronts in the images, and their direction of propagation that is oblique to the direction of the magnetic field. Here as well no temporal damping of the oscillations is evident. Figure~\ref{vf30v0_05xp01_2:fig}  shows the temporal evolution near the apex of the long magnetic loop (point $D=(0.01,-0.01,2.5)$) for the periodically driven upflow case. The shock-like structures are evident in the temporal evolution of the velocity components, the density, and the magnetic field exhibiting strong nonlinear steepening. Here as well, the temporal evolution of $|\Delta B|$  shows two dominant frequencies due to the combined effect of the driver and the slow mode wave in the loop. The structures are shock-like (more diffuse compared to shocks), since they travel less than a single scale height, and their amplitude is significant but not large compared to the sound speed and fast magnetosonic speed. Convergence test shows that numerical dissipation does not affect significantly the shock-like waves in the present calculation. The nonlinear steepening of the flow fronts in the loop is due to inflow amplitude comparable to the sound speed at the footpoint, and the  density stratification in long loops (see Figure~\ref{loop_s:fig}).

Finally, we summarize the results of the four cases discussed above using 3D isocontours snapshots of the perturbed density in Figure ~\ref{rh3d:fig}. The four cases discussed above are: (a) steady inflow in the short magnetic loop; (b) periodic inflow in the short magnetic loop; (c) steady inflow in the  long magnetic loop; (d) periodic inflow in the long magnetic loop. By comparing between the isocontour snapshots and animations the effects of the magnetic loop size and the types of inflow become apparent. A steady loop is formed following steady inflow and induced damped oscillations in the initial transitory stage (longitudinal and transverse)  in the short loop that lays entirely in the low-$\beta$ region of the AR. When the periodic flow is injected in such magnetic loop the density structure in the loop is dominated by the structure of the flow with small effects due to waves in the loop. In the long loop the steady state is not reached, and the density of the loop is variable throughout the evolution for both types of inflow. When the inflow is periodic the effects of the waves induced by the flow are significant in the long loop (for the magnitude of the flow in the present study) and the density structure exhibits the coupled fast magnetosonic and the slow wave shock-like structure in the high-$\beta$ region of the model AR.

\section{Discussion and Conclusions}

High resolution and high cadence observations indicate that some coronal loop oscillations are associated with flows along the loops and with jets at loops' footpoints. Whether propagating intensity disturbances seen in EUV imaging observations of the coronal loop structures are flows or waves is still under intensive debate. On one hand the propagating disturbances have been interpreted as signatures of slow magnetoacoustic waves since their propagating speeds are close to the sound speed in the corona, and 3 min or 5 min periodicities suggest their association with the leakage of the photospheric $p$-modes \citep[e.g.,][]{Nak00,DeM02}. Combined stereoscopic EUV and spectral observations provide further evidence for slow magnetosonic waves in coronal loops \citep[e.g.,][]{Mar09,MW09}. Observations of propagating disturbances in coronal plumes were interpreted as slow magnetosonic waves not only based on their speed but also on their amplitudes' dependence with height, found to be in agreement with the expected variation of the wave amplitude with height in gravitationally stratified plasma \citep{OND99}. On the other hand, some recent studies suggested that propagating EUV disturbances are most likely intermittent transient upflows because their association with persistent outflows at the loop base \citep[e.g.,][]{Har08b}, and their correlation with blueward asymmetries of line profiles using Hinode/EIS observations \citep{MD09,Tia11}.  However, these observational signatures  were found to be consistent with slow mode waves as well \citep{Verw10}. 

Our 3D MHD model results of flows and associated waves in hot coronal loops show the close connection between the two phenomena and suggest that both may be present simultaneously and produced by the same impulsive events at the footpoints (or corona/transition region interface) of coronal loops. In particular, we study the effects of localized inflow at the lower coronal boundary in a bi-polar magnetic field of hot AR coronal loops. The 3D MHD model reproduced the basic physical coupling between steady and periodic flows, waves, and loop density and magnetic structures in realistic AR geometry, with applications to coronal seismology, extending  previous studies that use more simplified models and help addressing the controversy of flows vs. waves discussed above. The main limitations of our AR model are the use of idealized  magnetic structure with dipole field, and gravitationally stratified density in the initial state. In addition the isothermal energy equation is used,  simplifying the calculations considerably at the expense of calculating the effects of heat conduction, heating and cooling of the material injected into the loops, and the possible heating of the loops by the energy released in the impulsive events. 

We investigate the effects of steadily injected inflow in short and long magnetic loops and the associated oscillations in the initial impulsive stage. We find that the inflow at the footpoints of coronal magnetic loops leads to formation of higher density than the surroundings loops due to the siphon effect in the curved bipolar magnetic structure. The inflows excite coupled longitudinal and transverse oscillations in the loops in the initial (impulsive) phase that later damp primarily through leakage as the loops reach the new steady state with background flow. The fundamental standing slow mode waves are seen to form in short and long loops in agreement with past observations and models discussed above. The fast magnetosonic waves induced by the inflow and nonlinearity are detected as well. In short loops with steady inflow in low-$\beta$ regions of the AR the properties of the oscillations (period, wavelength, damping rate), depend on loop length, and curvature of the magnetic structure in qualitative agreement with previous studies of waves in loops without background flow \citep{MO08,SO09,SeO10,SOS11,SSO11}. We also find that the flows produce impulsively nonlinear fast mode waves in the loops.

We investigate the effects of quasi-periodic flow injection in the localized region at lower coronal boundary, and find that driven undamped oscillations are produced in the magnetic loop of the bipolar AR's. As a result, the average density of the loops increases as well in this case. The period, amplitude, and damping rate of the driven oscillations are determined primarily by the properties of the driving periodic flows such as their period and amplitude, and the nonlinear interaction with the loop structure. In hot and long loops the quasi-periodic inflows with subsonic magnitude at the lower boundary can produce shock-like nonlinear compressive disturbances that propagate obliquely to the magnetic loop direction. 

We conclude that, a way to determine whether the observed loop oscillations result from slow waves or quasi-periodic flows is to evaluate the relation between the oscillation properties such as period, phase speed, phase relations, damping time, and the various loop parameters such as loop length, density, temperature, and magnetic field structure, since these relations are different for waves and for quasi-periodic flows as demonstrated in our study. The presence of the temporal damping of the oscillations on the time scale shorter than the decay time of the flow (as in the case of steady inflow) supports the wave interpretation of the observed oscillations. In the case of periodic inflow studied here the oscillations do not exhibit damping. We conclude that the excitation of slow-mode (and some transverse)  oscillations and flows observed with SUMER, TRACE, Hinode, and recently with SDO in coronal loops may result from the impulsive flow injection of plasma at the corona-chromosphere interface at the footpoints of the coronal AR loops.  Thus, our study suggests that flows and waves in coronal loops may result of the same impulsive events that drive these phenomena.

\acknowledgments We would like to acknowledge support by NASA grant NNX12AB34G. LO was also supported by NASA grants NNX08AV88G, NNX09AG10G, and NNX10AN10G. TJW was also supported by NASA grants NNX10AN10G and NNX08AE44G. We acknowledge the use of computer resources at NASA’s Ames Research Center advanced supercomputing facility.

\begin{thebibliography}{42}
\expandafter\ifx\csname natexlab\endcsname\relax\def\natexlab#1{#1}\fi

\bibitem[{{Aschwanden} {et~al.}(2002){Aschwanden}, {De Pontieu}, {Schrijver},
  \& {Title}}]{Asc02}
{Aschwanden}, M.~J., {De Pontieu}, B., {Schrijver}, C.~J., \& {Title}, A.~M.
  2002, \solphys, 206, 99

\bibitem[{{Berghmans} \& {Clette}(1999)}]{BC99}
{Berghmans}, D., \& {Clette}, F. 1999, \solphys, 186, 207

\bibitem[{{De Moortel} {et~al.}(2002){De Moortel}, {Hood}, \&
  {Ireland}}]{DeM02}
{De Moortel}, I., {Hood}, A.~W., \& {Ireland}, J. 2002, \aap, 381, 311

\bibitem[{{De Pontieu} \& {McIntosh}(2010)}]{DeP10}
{De Pontieu}, B., \& {McIntosh}, S.~W. 2010, \apj, 722, 1013

\bibitem[{{Hara} {et~al.}(2008){Hara}, {Watanabe}, {Harra}, {Culhane}, {Young},
  {Mariska}, \& {Doschek}}]{Har08b}
{Hara}, H., {Watanabe}, T., {Harra}, L.~K., {Culhane}, J.~L., {Young}, P.~R.,
  {Mariska}, J.~T., \& {Doschek}, G.~A. 2008, \apjl, 678, L67

\bibitem[{{Kliem} {et~al.}(2002){Kliem}, {Dammasch}, {Curdt}, \&
  {Wilhelm}}]{Kli02}
{Kliem}, B., {Dammasch}, I.~E., {Curdt}, W., \& {Wilhelm}, K. 2002, \apjl, 568,
  L61

\bibitem[{{Klimchuk}(2009)}]{kli09}
{Klimchuk}, J.~A. 2009, in Astronomical Society of the Pacific Conference
  Series, Vol. 415, The Second Hinode Science Meeting: Beyond Discovery-Toward
  Understanding, ed. {B.~Lites, M.~Cheung, T.~Magara, J.~Mariska, \&
  K.~Reeves}, 221--+

\bibitem[{{Lionello} {et~al.}(1998){Lionello}, {Velli}, {Einaudi}, \&
  {Mikic}}]{Lio98}
{Lionello}, R., {Velli}, M., {Einaudi}, G., \& {Mikic}, Z. 1998, \apj, 494, 840

\bibitem[{{Marsh} \& {Walsh}(2009)}]{MW09}
{Marsh}, M.~S., \& {Walsh}, R.~W. 2009, \apjl, 706, L76

\bibitem[{{Marsh} {et~al.}(2009){Marsh}, {Walsh}, \& {Plunkett}}]{Mar09}
{Marsh}, M.~S., {Walsh}, R.~W., \& {Plunkett}, S. 2009, \apj, 697, 1674

\bibitem[{{McIntosh} \& {De Pontieu}(2009)}]{MD09}
{McIntosh}, S.~W., \& {De Pontieu}, B. 2009, \apj, 707, 524

\bibitem[{{McLaughlin} \& {Ofman}(2008)}]{MO08}
{McLaughlin}, J.~A., \& {Ofman}, L. 2008, \apj, 682, 1338

\bibitem[{{Nakariakov} \& {Roberts}(1995)}]{Nak95}
{Nakariakov}, V.~M., \& {Roberts}, B. 1995, \solphys, 159, 213

\bibitem[{{Nakariakov} \& {Verwichte}(2005)}]{NV05}
{Nakariakov}, V.~M., \& {Verwichte}, E. 2005, Living Reviews in Solar Physics,
  2, 3

\bibitem[{{Nakariakov} {et~al.}(2000){Nakariakov}, {Verwichte}, {Berghmans}, \&
  {Robbrecht}}]{Nak00}
{Nakariakov}, V.~M., {Verwichte}, E., {Berghmans}, D., \& {Robbrecht}, E. 2000,
  \aap, 362, 1151

\bibitem[{{Nishizuka} \& {Hara}(2011)}]{Nis11}
{Nishizuka}, N., \& {Hara}, H. 2011, \apjl, 737, L43+

\bibitem[{{Ofman}(2007)}]{Ofm07}
{Ofman}, L. 2007, \apj, 655, 1134

\bibitem[{{Ofman}(2009)}]{Ofm09}
---. 2009, \apj, 694, 502

\bibitem[{{Ofman} {et~al.}(1999){Ofman}, {Nakariakov}, \& {Deforest}}]{OND99}
{Ofman}, L., {Nakariakov}, V.~M., \& {Deforest}, C.~E. 1999, \apj, 514, 441

\bibitem[{{Ofman} {et~al.}(2000){Ofman}, {Nakariakov}, \& {Sehgal}}]{ONS00}
{Ofman}, L., {Nakariakov}, V.~M., \& {Sehgal}, N. 2000, \apj, 533, 1071

\bibitem[{{Ofman} \& Selwa(2009)}]{OS09}
{Ofman}, L., \& Selwa, M. 2009, in {IAU Symposium 257: Universal Heliophysical
  Processes}, ed. A.~{Nindos} \& etal. (New York, NY: Cambridge University
  Press)

\bibitem[{{Ofman} \& {Thompson}(2002)}]{OT02}
{Ofman}, L., \& {Thompson}, B.~J. 2002, \apj, 574, 440

\bibitem[{{Ofman} \& {Wang}(2008)}]{OW08}
{Ofman}, L., \& {Wang}, T.~J. 2008, \aap, 482, L9

\bibitem[{{Roberts} {et~al.}(1984){Roberts}, {Edwin}, \& {Benz}}]{REB84}
{Roberts}, B., {Edwin}, P.~M., \& {Benz}, A.~O. 1984, \apj, 279, 857

\bibitem[{{Schmidt} \& {Ofman}(2011)}]{SmO11}
{Schmidt}, J.~M., \& {Ofman}, L. 2011, \apj, 739, 75

\bibitem[{{Selwa} {et~al.}(2005){Selwa}, {Murawski}, \& {Solanki}}]{Sel05b}
{Selwa}, M., {Murawski}, K., \& {Solanki}, S.~K. 2005, \aap, 436, 701

\bibitem[{{Selwa} \& {Ofman}(2009)}]{SO09}
{Selwa}, M., \& {Ofman}, L. 2009, Annales Geophysicae, 27, 3899

\bibitem[{{Selwa} \& {Ofman}(2010)}]{SeO10}
---. 2010, \apj, 714, 170

\bibitem[{{Selwa} {et~al.}(2007){Selwa}, {Ofman}, \& {Murawski}}]{Sel07}
{Selwa}, M., {Ofman}, L., \& {Murawski}, K. 2007, \apjl, 668, L83

\bibitem[{{Selwa} {et~al.}(2011{\natexlab{a}}){Selwa}, {Ofman}, \&
  {Solanki}}]{SOS11}
{Selwa}, M., {Ofman}, L., \& {Solanki}, S.~K. 2011{\natexlab{a}}, \apj, 726, 42

\bibitem[{{Selwa} {et~al.}(2011{\natexlab{b}}){Selwa}, {Solanki}, \&
  {Ofman}}]{SSO11}
{Selwa}, M., {Solanki}, S.~K., \& {Ofman}, L. 2011{\natexlab{b}}, \apj, 728, 87

\bibitem[{{Svestka}(1994)}]{Sve94}
{Svestka}, Z. 1994, \solphys, 152, 505

\bibitem[{{Tian} {et~al.}(2011){Tian}, {McIntosh}, \& {De Pontieu}}]{Tia11}
{Tian}, H., {McIntosh}, S.~W., \& {De Pontieu}, B. 2011, \apjl, 727, L37+

\bibitem[{{Verwichte} {et~al.}(2010){Verwichte}, {Marsh}, {Foullon}, {Van
  Doorsselaere}, {De Moortel}, {Hood}, \& {Nakariakov}}]{Verw10}
{Verwichte}, E., {Marsh}, M., {Foullon}, C., {Van Doorsselaere}, T., {De
  Moortel}, I., {Hood}, A.~W., \& {Nakariakov}, V.~M. 2010, \apjl, 724, L194

\bibitem[{{Wang}(2011)}]{Wan11b}
{Wang}, T. 2011, \ssr, 158, 397

\bibitem[{{Wang} {et~al.}(2010){Wang}, {Ofman}, \& {Davila}}]{Wan10}
{Wang}, T.~J., {Ofman}, L., \& {Davila}, J. 2010, in Astronomical Society of
  the Pacific Conference Series, Vol. 455, 4th Hinode Science Meeting: Unsolved
  Problems and Recent Insights, ed. {Luis R. Bellot Rubio}, 227

\bibitem[{{Wang} {et~al.}(2009){Wang}, {Ofman}, {Davila}, \& {Mariska}}]{Wan09}
{Wang}, T.~J., {Ofman}, L., {Davila}, J.~M., \& {Mariska}, J.~T. 2009, \aap,
  503, L25

\bibitem[{{Wang} {et~al.}(2002){Wang}, {Solanki}, {Curdt}, {Innes}, \&
  {Dammasch}}]{Wan02}
{Wang}, T.~J., {Solanki}, S.~K., {Curdt}, W., {Innes}, D.~E., \& {Dammasch},
  I.~E. 2002, \apjl, 574, L101

\bibitem[{{Wang} {et~al.}(2003{\natexlab{a}}){Wang}, {Solanki}, {Curdt},
  {Innes}, {Dammasch}, \& {Kliem}}]{Wan03b}
{Wang}, T.~J., {Solanki}, S.~K., {Curdt}, W., {Innes}, D.~E., {Dammasch},
  I.~E., \& {Kliem}, B. 2003{\natexlab{a}}, \aap, 406, 1105

\bibitem[{{Wang} {et~al.}(2005){Wang}, {Solanki}, {Innes}, \& {Curdt}}]{Wan05}
{Wang}, T.~J., {Solanki}, S.~K., {Innes}, D.~E., \& {Curdt}, W. 2005, \aap,
  435, 753

\bibitem[{{Wang} {et~al.}(2003{\natexlab{b}}){Wang}, {Solanki}, {Innes},
  {Curdt}, \& {Marsch}}]{Wan03a}
{Wang}, T.~J., {Solanki}, S.~K., {Innes}, D.~E., {Curdt}, W., \& {Marsch}, E.
  2003{\natexlab{b}}, \aap, 402, L17

\bibitem[{{Wang} {et~al.}(2008){Wang}, {Solanki}, \& {Selwa}}]{Wan08}
{Wang}, T.~J., {Solanki}, S.~K., \& {Selwa}, M. 2008, \aap, 489, 1307

\end{thebibliography}
\end{document}